\documentclass[letter]{aa}
\usepackage{graphicx}
\usepackage{epsfig}
\usepackage{latexsym}
\usepackage{txfonts}
\begin{document}
\renewcommand{\labelitemi}{-}
\title{Exploring the connection between coronal and footpoint sources in a thin-thick target solar flare model}
\author{Marina Battaglia
  \and Arnold O. Benz}
\institute{Institute of Astronomy, ETH Zurich, 8092 Zurich, Switzerland}
\date{Received /Accepted}

\abstract
{Hard X-ray emission of coronal sources in solar flares has been observed and studied
  since its discovery in Yohkoh observations. Several models have been
  proposed to explain the physical mechanisms causing this emission and the relations between those sources and simultaneously observed footpoint sources.}
{We investigate and test one of the models (intermediate thin-thick target model)
  developed on the basis of Yohkoh observations. The model makes precise
  predictions on the shape of coronal and footpoint spectra and the relations between them, that can be
  tested with new instruments such as RHESSI.}
{RHESSI observations of well observed events are studied in imaging and
  spectroscopy and compared to the predictions from the intermediate
  thin-thick target model.}
{The results indicate that such a simple model cannot account for the observed relations between the non-thermal spectra of coronal and footpoint sources. Including non-collisional energy loss of the electrons in the flare loop due to an electric field can solve most of the inconsistencies. }
{}

\titlerunning{Relations between X-ray sources in solar flares}
\authorrunning{Marina Battaglia \& Arnold O. Benz}

\maketitle


\section{Introduction} \label{Introduction}
How well do we understand the physics behind solar flares? Solar flares
release a large portion of their energy within seconds to minutes. A significant part of
this energy goes into accelerated electrons and ions which then precipitate to
the chromosphere along the field lines of a magnetic loop. The chromosphere
acts as a thick target on the precipitating particles, leading to
characteristic hard X-ray (HXR) emission. The first observations of such footpoint sources
were made by  Hoyng et al. (\cite{Ho81}). Masuda et al. (\cite{Ma94}) found
another HXR source near the top of flare loops in Yohkoh observations. Feldman
et al. (\cite{Fe94}) analyzed such coronal sources with Yohkoh BCS and SXT. Due to
the normally low coronal densities one would expect a thin target emission
from this coronal source. However, the observations by Feldman et
al. (\cite{Fe94}) yield quite high column densities
($\mathrm{10^{20}cm^{-2}<N<7\cdot 10^{20}cm^{-2}}$) in the coronal source
that would act as a partly thick target on electrons below a certain energy. 
On the basis of the observations by Masuda et al. (\cite{Ma94}) and Feldman et
al. (\cite{Fe94}), Wheatland and Melrose (\cite{We95}) developed a simple
model that has been used and investigated further (e.g. Metcalf $\&$ Alexander \cite{Al99}, Fletcher $\&$ Martens~\cite{Fl98} ). The model consists of four
basic elements; a particle accelerator on top of a magnetic loop, a coronal source visible in SXR and
HXR, collision-less propagation of particles along the magnetic loop and
HXR-footpoints in the chromosphere. The coronal source acts as an intermediate
thin-thick target on electrons depending on energy (thick target for lower
energetic electrons, thin target on higher energies). We will therefore refer
to this model as intermediate thin-thick target, or ITTT model. 

For large enough column depths or steep enough electron spectra, almost all electrons would be stopped in the loop and no footpoints could be observed. Such events were observed with Yohkoh and later with RHESSI. Veronig \& Brown~(\cite{Ve04}) analyzed flares with faint footpoint emission but a dense loop acting as a thick target on electrons of energies up to 60 keV. 

Due to its high spectral and spatial resolution, RHESSI (Lin et
al. \cite{Li02}) provides the means to perform imaging spectroscopy on events
with multiple sources. Battaglia $\&$ Benz (\cite{Ba06}) showed that it is
possible to study the non-thermal X-ray emission of coronal sources and
footpoints and the relations between them in detail. 

Therefore, the  ITTT model and its predictions can be tested using RHESSI
data of well observed events. It is the simplest model that can explain the Yohkoh data. Can it also explain RHESSI data?  
\section{Theoretical model} \label{Model}
\begin{figure}
\resizebox{\hsize}{7cm}{\includegraphics{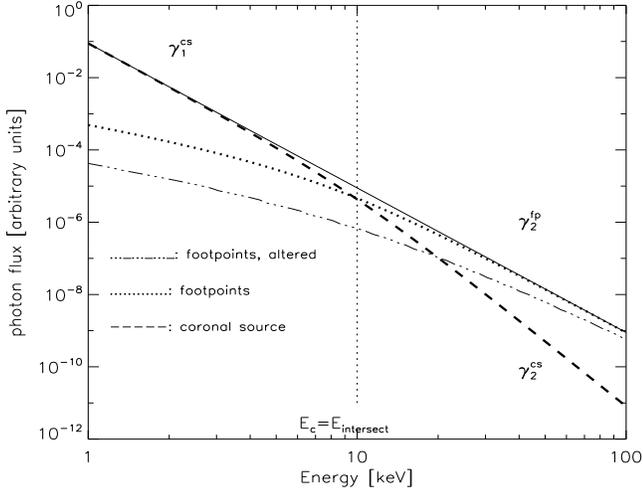}}
\caption {Schematic photon spectra for electron distribution with power-law spectral index $\delta=5$ and
  critical energy $E_c=10$ keV. The dashed line gives the spectrum of the
  coronal source with the spectral indices $\gamma_{1}^{cs}=\gamma_2^{fp}=\delta-1$ and
  $\gamma_{2}^{cs}=\delta+1$. The dotted line indicates the combined spectrum of the two footpoints. The
  footpoint spectrum is a power-law with spectral index $\gamma_2^{fp}=\delta-1$ at
  high energies. The solid line is the total of coronal source and footpoints (pure thick target). The dash-dotted line gives the footpoint spectrum from an electron population that has been altered in the loop (see Discussion).}
\label{photspec}
\end{figure}

The ITTT model features a dense coronal source into which a beam of electrons with
a power-law energy distribution is injected. Some high energy electrons then leave the dense
region, precipitating down to the chromosphere. The coronal region acts as a
thick target on particles with energy lower than a critical energy $E_c$ and
as thin target on electrons with energy larger than the critical energy. This
results in a characteristic HXR spectrum as well as SXR emission due to
collisional heating of the coronal region. The altered electron beam reaches
the chromosphere, causing thick target emission in the footpoints of the
magnetic loop. The predicted photon spectra are shown in
Fig.~\ref{photspec}. The coronal spectrum consists of two power-law
components with a break at $E_c$. The footpoint spectrum has a break at the same energy
as the coronal spectrum and is a power-law at high
energies.

The model predicts the following properties:
\begin{enumerate}

\item The coronal and footpoint spectra intersect at the critical energy ($E_c=E_{intersect}$).
\item There is a break at $E_{c}$ in the individual spectra of coronal source and footpoints, but not in the sum of the two spectra.
\item The spectral indices below and above the break in the coronal source have a
  difference of $\gamma_{2}^{cs}-\gamma_{1}^{cs}=2$. 
\item The spectral index of the coronal source below the break is equal to the
  spectral index of the footpoints above the break ($\gamma_1^{cs}=\gamma_2^{fp}$).
\item The difference between the spectral indices of the coronal source and
  the footpoints at high energies is $\gamma_{2}^{cs}-\gamma_2^{fp}=2$.
\item Coronal source and footpoints have the same intensity at $E_c$.
\item SXR and HXR emission in the coronal source are spatially
  coincident. 
\end{enumerate}

\section{Observations}
\subsection{Event selection and spectral analysis}
We analyzed the five events described in Battaglia $\&$ Benz (\cite{Ba06}) in an
interval of 3 RHESSI spin-periods (about 12 s) around the time of maximum
HXR-emission. The events are listed in Table~\ref{eventlist}. 

\setlength{\tabcolsep}{1.2mm}
\begin{table}
\caption{Eventlist. The time indicates the peak time in the 50 - 100 keV energy band.}
\begin{center}
\begin{tabular}{lccr}  
\hline \hline
No.& Date & Time & GOES class \\
\hline
1& 4-Dec-2002 & 22:47:02 & M2.7          \\
2& 24-Oct-2003&02:48:42 & M7.7       \\
3& 1-Nov-2003 &  22:33:14& M3.3             \\
4& 13-Jul-2005 & 14:14:30 & M5.1       \\
5& 30-Jul-2005 & 06:32:06 & X1.3          \\
\hline
\end{tabular}
\end{center}
 \label{eventlist}
\end{table}

Imaging spectroscopy applying the Clean algorithm for image
reproduction was used to produce spectra of coronal sources and
footpoints. Reasons for favoring Clean over Pixon are discussed in Battaglia \& Benz \cite{Ba06}. To compare the observations with the model, the
footpoints where treated as one region and only a total spectrum over both 
footpoints has been computed.
\begin{figure*}
\resizebox{\hsize}{9cm}{\includegraphics{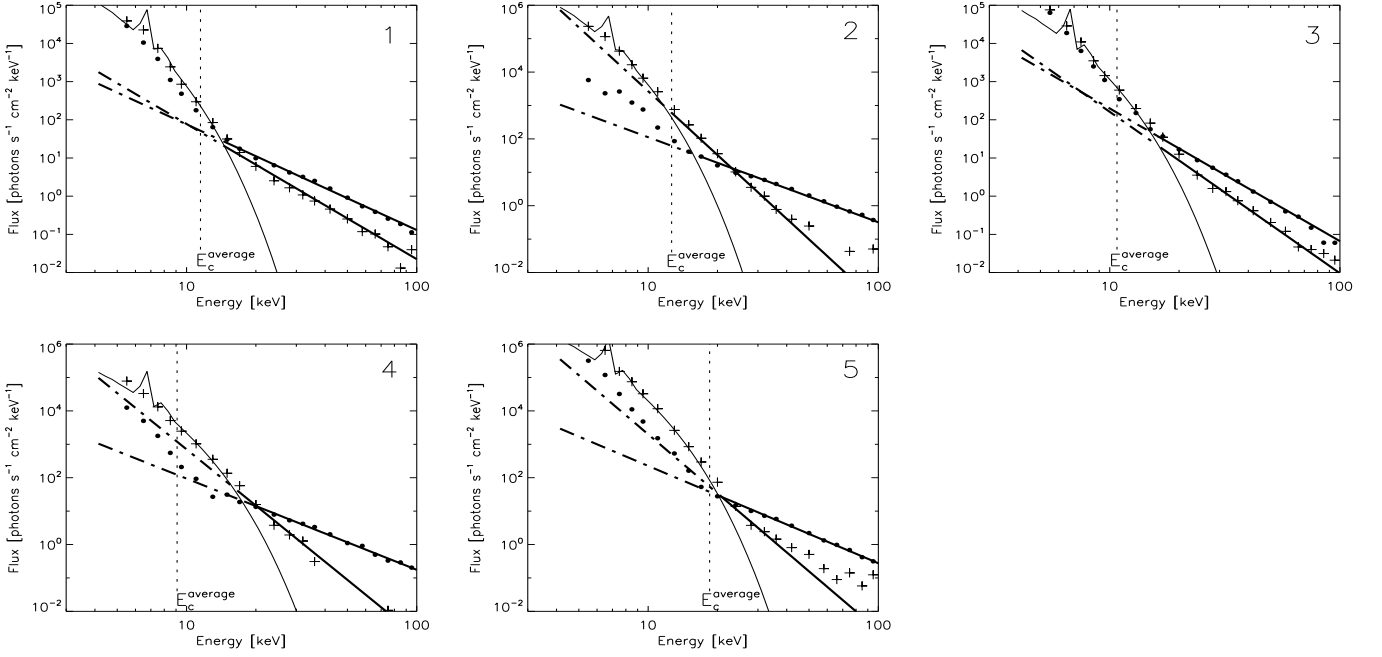}}
\caption {Coronal (\textit{crosses}) and footpoint (\textit{dots}) spectra 
of the five analyzed events. The thin solid line indicates the thermal fit to
the coronal source. The thick solid lines give the power-law fits to coronal
source and footpoints. Their extension to low energies (not observed) is given
as a dot-dashed line. The dotted line indicates the average critical energy.}
\label{spectra}
\end{figure*}
The coronal sources have been fitted with a thermal component and a non-thermal
power-law, the footpoints only with a non-thermal component. Varying the starting fit parameters and the fit energy range provided a validation for the stability of the fit and an estimate of the range of spectral index values. The spectra and fits are shown in Fig.~\ref{spectra}.
\subsection{Density and critical energy}
In the ITTT model, the critical electron energy for the transition between
thick and thin target for a given column density $N_0$ is given by equating the source size to half of the mean free path,
\begin{equation}\label{ecrit}
E_c=\sqrt{2KN_0}
\end{equation} with the constant $K=2\pi e^4\ln{\Lambda}$. The Coulomb
logarithm $\mathrm{ln{\Lambda}}$ is about 20 for the electron densities and
temperatures in the presented sample.  
The column depth the electron beam passes in the coronal source was computed
using RHESSI and GOES data, following the treatment in Wheatland
\& Melrose (\cite{We95}). For a source volume $V$, a total source
diameter of $V^{1/3}$ is assumed. An electron beam injected in the middle of the source
would then travel half this distance, giving a column depth of: 
\begin{equation}\label{no}
N_o=\frac{n_eV^{1/3}}{2}=\sqrt{EM}\cdot \frac{A^{-1/4}}{2}
\end{equation}
where the electron density $n_e=\sqrt{EM/V}$ using the observed emission measure EM and the
volume $V=A^{3/2}$. The area A was measured from the 50$\%$ contour of the coronal
source in RHESSI Clean images at energy 10 - 12 keV, taking into
account the Clean-Beam size and effects of the pixel size on the contour determination.

The emission measure was computed from spectral fits to RHESSI full sun
spectra and the spectrum of the coronal source only. As noted by
Battaglia $\&$ Benz (\cite{Ba06}), the thermal footpoint emission generally is very faint in RHESSI
observations. Therefore, the thermal emission measured
in full sun spectra is predominantly coronal emission. However, the spectral fittings deviate slightly, the temperature being lower and the emission measure higher for the imaging spectroscopy fit. We therefore use both as a range of possible column densities.

Additionally, we computed the GOES emission measure and temperature
for comparison. The range of all  RHESSI and GOES emission measures then gives an estimate of the accuracy  for
the column densities and critical energies (Table \ref{results}). 
\subsection{Position of coronal source at different energies}
The ITTT model predicts spatial coincidence of the coronal emission for all energies. To check for this, Clean images at different energies where made and the centroid of the 50\% emission of the coronal source was computed in order to get the position. At energies higher than 20~keV, the footpoint emission starts to dominate, making an accurate determination of the coronal source-position difficult. We determined the centroid positions for the energy range 6-12 keV (thermal) and 16-22 keV (partially non-thermal).

\begin{figure}
\resizebox{\hsize}{!}{\includegraphics{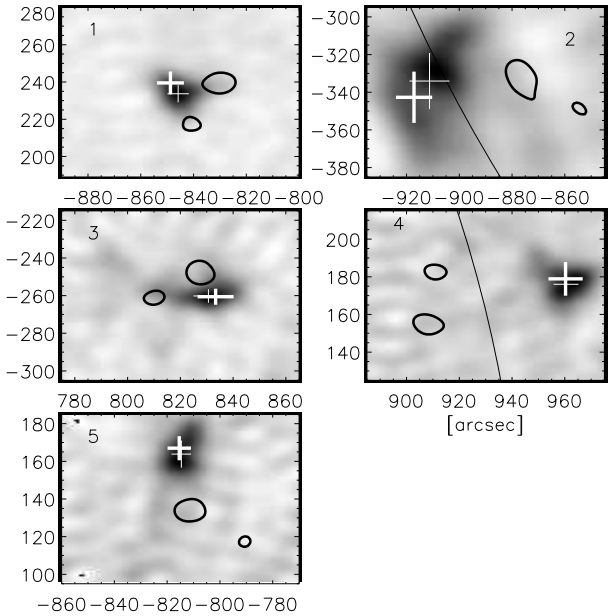}}
\caption {Images of all events at energies 10 - 12 keV. The 50\% contours of the maximum of the footpoints are given (black). The white crosses indicate the position of the 50\% centroid at energies 6 - 12 keV (thin) and 16 - 22 keV (thick) with error bars.}
\label{morph}
\end{figure}
\section{Results}
The main numerical results are given in Table~\ref{results}.
We find column densities between $2.1\cdot 10^{19}<N_0<1.2\cdot 10^{20}$ cm$^{-2}$.
This yields a range of $E_c$ from 7.7 keV - 21 keV,
significantly lower than the values calculated by Wheatland \&
Melrose for Yohkoh events (15 keV$< E_c <$ 40 keV).
\setlength{\tabcolsep}{1.2mm}
\begin{table*}
\caption{Summary of observed parameters to be compared with the theory. The
  power-law spectrum of the coronal source can only be observed at energies
  higher than the thermal emission, therefore no value for the part below the
  break could be determined.}
\begin{center}
\begin{tabular}{lccccc}   
\hline \hline
Event& 1 & 2 & 3 & 4 & 5 \\
\hline
$\gamma_{1}^{cs}$ & - &(6.3$\pm$0.1)&-&-&-    \\
$\gamma_{2}^{cs}$ &3.5$\pm$0.7&6.3$\pm$0.1&4.2$\pm$0.4&5.6$\pm$0.2&5.9$\pm$1.0 \\
$\gamma_2^{fp}$   &2.8$\pm$0.2&2.5$\pm$0.1&3.5$\pm$0.1&2.7$\pm$0.2&2.9$\pm$0.1 \\
$\gamma_{2}^{cs}-\gamma_2^{fp}$&0.7$\pm$0.7&3.8$\pm$0.1&0.7$\pm$0.4&2.9$\pm$0.3&3$\pm$ 1.0 \\
$N_0$ $\mathrm{(cm^{-2}})$&(2.5-4.9)$\cdot 10^{19}$&(2.8-6.1)$\cdot
10^{19}$&(2.1-4.3)$\cdot 10^{19}$&(2.6-3.0)$\cdot 10^{19}$&(7.1-12)$\cdot
10^{19}$ \\
$E_c$ (keV)&9.5-13.5&10.3-15&8.9-12.6&7.7-10.5&16.2-20.8 \\
$E_{intersect}$ (keV)&10.3&23.1&7.6&20.2&20.9 \\
\hline
\end{tabular}
\end{center}
 \label{results}
\end{table*}
The comparison of the data with the predictions from the ITTT model is not always possible due to the presence of the thermal component. Especially, the estimated critical energy is within the range dominated by the thermal component in all events. 
Comparing the data with the predictions from the ITTT model
 gives the following results, following the numbering in Sect.~\ref{Model}:
\begin{enumerate}

\item The power-law fits of the non-thermal coronal and footpoint spectra intersect at
  energies $E_{intersect}$= 7.6 - 23.1 keV. In 3 events, this energy is within or just outside the
  estimate for the critical energy derived from eqns. (\ref{ecrit}) and (\ref{no}). Therefore $E_c \approx E_{intersect}$. In
  two events (no. 2 and 4), $E_{intersect}$ is 
  significantly higher than the computed $E_c$ ($E_{intersect} > E_c$). 
\item In the events with  $E_c \approx E_{intersect}$ the estimated break energies and intersection energies are in the energy range
  dominated by the thermal emission of the coronal source. A break in the coronal
  spectrum can therefore not be detected, but might still exist. In events nos. 2 and 4, where $E_{intersect} > E_c$, the intersection point of the power-law spectra is observed, but
  without a break, contradicting the prediction. 
\item In event 2, the intersection energy is well outside the thermal emission but $\gamma_2^{cs}-\gamma_1^{cs}\neq 2$ in the coronal source, contrary to the prediction.
\item In event 2, $\gamma_1^{cs}>\gamma_2^{fp}$, not equal as expected from the model. 
\item $\gamma_2^{cs}-\gamma_2^{fp} > 2$ in three events and $\gamma_2^{cs}-\gamma_2^{fp} < 2$ in two events.
\item In three events, $E_c$ and $E_{intersect}$ of the non-thermal fits are within the thermal emission of the coronal source. In event 2, $E_{intersect}$ is observed with the coronal source being brighter than the footpoints at $E_c$.
\item No significant positional differences between energies 6 - 12 keV and
  16 - 22 keV have been found (see Fig.~\ref{morph}). Due to the increasing footpoint brightness at
  higher energies, an accurate determination of the coronal source centroid above 22 keV is not possible.

\end{enumerate}
\subsection{Energy input into the corona}
Wheatland $\&$ Melrose (\cite{We95}) proposed that the collisional energy necessary to produce the non-thermal bremsstrahlung emission in the coronal source is sufficient to account for  the observed heating of the coronal source. They referred to standard flare
values (flux, duration, electron spectral index) to estimate the non-thermal energies. With the observations presented
here the energy input into the individual coronal source can be determined.  

We compute the change in thermal energy $\Delta E_{th}$ in the coronal source between two times in the flare. The thermal energy is calculated according to $E_{therm}=3k_BT\sqrt{EM\cdot V}$ with a filling factor of unity. 

The non-thermal energy input in the coronal source is computed from the non-thermal coronal spectrum. As cutoff energy, we used the intersection point of the thermal and non-thermal emission. Saint-Hilaire \& Benz (\cite{Sh05}) found that the average relation between photon-turnover and electron-cutoff is $E_{cutoff}^{electron}=E_{turnover}^{photon}\cdot 1.7$. This gives an upper limit for the electron cutoff energy and therefore a lower limit for the total energy input. 

Comparing the two lower limits, we find that the average non-thermal energy input is at least of the same order as the change in thermal energy i.e. $\Delta E_{non\_th}\ge \Delta E_{th}$ in agreement with the prediction of the ITTT model. For a lower electron cut-off energy, the non-thermal energy input could be up to an order of magnitude higher.

\section{Discussion}

The ITTT model as presented by Wheatland \& Melrose (\cite{We95}) predicts a critical electron energy $E_c$ marking the transition between thin- and thick target emission. This energy is likewise the intersection energy $E_{intersect}$ between the non-thermal spectra of the coronal and footpoint sources. How can it then be explained in the frame of the ITTT model that in events 2 and 4 the estimated $E_c$ is significantly lower than the observed $E_{intersect}$? 

$E_c$ as calculated here may be a lower limit. Including collisional deflections of the electrons within the coronal target rather than just energy-loss, the electron paths would become longer and the critical energy for thin target emission higher. This would increase $E_c$ and could bring it up towards the observed $E_{intersect}$. In that case however, one would expect to observe a break at $E_{intersect}$. This is not the case. Further, the difference $\gamma_{cs}^2-\gamma_{fp}^2$ is higher than 2 in both of these events, making this explanation even less likely. 

An attractive possibility to interpret both, the absence of a break and the larger difference in spectral index is non-collisional energy loss of the electrons in the loop due to the electric field caused by the return current. This would result in a  harder and fainter footpoint spectrum, shifting $E_{intersect}$ to energies higher than $E_c$. The expected footpoint spectrum in the case of electrons loosing 15 keV of energy in the loop is indicated in Fig.~\ref{photspec}. 

The remaining discrepancy then is the difference in spectral index $\gamma_{cs}^2-\gamma_{fp}^2$ being smaller than two in events 1 and 3. Such a difference could result where the targets cannot be clearly separated into thin and thick, for example if the sources are close as is the case in those events. Part of the emission selected for the footpoints could originate from the coronal source, giving a mixed thin-thick target spectrum for the footpoints. More observations are necessary to confirm this explanation.

An estimate of the non-thermal energy input into the coronal source compared to the change of the thermal energy content in the coronal source shows that collisions as manifest in hard X-ray emission may account for the observed heating of the coronal source. Depending on the electron cut-off energy, the amount of energy deposited in the target might even exceed the observed thermal energy. However, we did not consider cooling of the coronal source due to thermal conduction or radiation. Such effects could lead to a thermal energy input larger than observed.

\section{Conclusions}
We showed that a simple thin-thick target model as the one proposed by Wheatland \& Melrose (\cite{We95}) cannot explain all observations made with RHESSI. Modifications of the model are necessary, the simplest being the consideration of the electric field due to the return current. In the absence of a balancing ion flux, such a return current is predicted from basic physics in view of footpoint sources.

\begin{acknowledgements}
RHESSI data analysis at ETH Z\"urich is supported by ETH grant TH-1/04-2 and the Swiss National Science Foundation (grant 20-105366). We thank Paolo Grigis for helpful comments and discussions.
\end{acknowledgements}

\end{document}